\documentclass{webofc}
\usepackage[varg]{txfonts}   % Web of Conferences font
\usepackage{amsmath, amssymb}
\usepackage{subcaption}
\usepackage[usenames,dvipsnames]{color}
\usepackage{float}
\begin{document}
\title{Baryon Number Transport, Strangeness Conservation and $\Omega$-hadron Correlations}

\author{\firstname{Xiatong} \lastname{Wu}\inst{1}\fnsep\thanks{\email{maxwoo@physics.ucla.edu}} \and
        \firstname{Weijie} \lastname{Dong}\inst{2}\fnsep\thanks{\email{wjdong19@fudan.edu.cn}} \and
        \firstname{Xiaozhou} \lastname{Yu}\inst{2}\fnsep\thanks{\email{xzyu18@fudan.edu.cn}} \and
        \firstname{Hui} \lastname{Li}\inst{2} \and
        \firstname{Gang} \lastname{Wang}\inst{1} \and
        \firstname{Huan Zhong} \lastname{Huang}\inst{1,2} \and
        \firstname{Zi-Wei} \lastname{Lin}\inst{3}
        % etc.
}

\institute{Department of Physics and Astronomy, University of California, Los Angeles, CA 90095, USA 
\and
           Key Laboratory of Nuclear Physics and Ion-beam Application, Fudan University, Shanghai, China 
\and
           Department of Physics, East Carolina University, Greenville, NC 27858, USA
          }

\abstract{%
Although strange quarks are produced in $s\bar{s}$ pairs, the ratio of $\Omega^{-}$ to ${\bar{\Omega}}^{+}$ is greater than one in heavy-ion collisions at lower RHIC energies. Thus the produced $\Omega$ hyperons must carry net baryon quantum numbers from the colliding nuclei. We present results of $K$-$\Omega$ correlations from AMPT model simulations of Au+Au collisions at $\sqrt{s_{NN}}$ = 14.6 GeV, to probe dynamics for baryon number transport to mid-rapidities at this beam energy. We use both the default and string-melting versions to illustrate how hadronization schemes of quark coalescence and string fragmentations could leave imprints on such correlations. Implications on the measurements of these correlations with the STAR experiment at RHIC will also be discussed. 
}
\maketitle
\section{Introduction}
\label{intro}
In heavy-ion collisions at $\sqrt{s_{NN}} < 50$ GeV, the anti-hyperon to hyperon ratios are significantly below one \cite{besi}, which implies that net hyperons must carry the net baryon number from the incident nuclei. The $\it{u}$ or $\it{d}$ quarks inside $\Lambda$'s and $\Xi$'s may originate from the colliding nuclei, and thus carry a portion of the net baryon number. However, there must be a mechanism for net baryon numbers to transport to $\Omega$'s without the delivery of $\it{u}$ or $\it{d}$ quarks, since the three $\it{s}$ quarks in $\Omega$ are produced in pairs. 

$\Omega$ production can carry dynamical information from strangeness conservation (SC), baryon number conservation (BNC), and baryon number transport (BNT). A primordial $\Omega^{-}$ is either produced along with three kaons and no anti-baryon (SC+BNT), or with one anti-baryon and some kaons (SC+BNC). In the former (SC+BNT), the $\it{s}$ quarks in the kaons may combine with $\it{u}$ or $\it{d}$ quarks that originate from the colliding nuclei, and thus the $\Omega^{-}$ can effectively obtain baryon numbers from them. This process may also be sensitive to a possible gluonic junction mechanism of baryon number transport \cite{junction}, where the baryon number is carried by the junction rather than by the valence quarks. The $\Omega^{-}$ arises from the hadronization of the junction by obtaining three $s$ quarks, whereas the kaons can be viewed as the "leading mesons" \cite{huang_proc}. In the latter (SC+BNC), all quarks are pair-produced and there is no net baryon number associated with the $\Omega^{-}$. These two scenarios can be characterized by $\Delta N_K$ and $\Delta N_{\bar{B}}$, the differences in the numbers of $K$'s ($K^{+}$ and $K^{0}$) and anti-baryons, respectively, between events with an $\Omega^{-}$ and events without any, as shown in Table~\ref{scenarios}. These values can be counted precisely in models with different hadronization schemes, and provide baselines  for more accessible observables in experiments, such as $K$-$\Omega$ and $\bar{B}$-$\Omega$ correlations.

\begin{table}[htb]
\centering
\caption{Expected differences in $N_K$ and $N_{\bar{B}}$ between events with $\Omega^{-}$ and events without any.}
\label{scenarios}       % Give a unique label
% For LaTeX tables you can use
\begin{tabular}{lll}
\hline
   & $\Delta N_{K}$ & $\Delta N_{\bar{B}}$  \\\hline
Scenario 1 (SC+BNT) & 3 & 0\\
Scenario 2 (SC+BNC) & 1-3 & 1\\\hline
\end{tabular}
\end{table}

\section{Model and Energy Selection}
\label{sec-1}
We choose A Multi-Phase Transport (AMPT) model \cite{AMPT} (non-public version v1.25t4cu/v2.25t5cu) to perform simulations for this study, with both the default  and  string-melting versions. The former includes only the minijet partons in the parton cascade phase and uses the Lund string fragmentation model to convert strings to hadrons. In the latter, both excited strings and minijet partons participate in parton cascade, and a quark coalescence model is used for hadronization. Comparison between these two modes may reveal the difference in the final-state manifestation of the two hadronization schemes. This version of AMPT also satisfies the quantum number (strangeness, electric charge and baryon number) conservation event by event, facilitating the study of the interplay of these quantum numbers in the context of the $\Omega$ production. 

We have generated about 150 million (50 million default + 100 million string-melting) minimum bias events of Au+Au at $\sqrt{s_{NN}} = 14.6$ GeV. In the majority of the events, at most one $\Omega^{-}$($\bar{\Omega}^{+}$) is produced.  Table \ref{diff:a} shows that the difference in the average number of $s\bar{s}$ pairs between events with and without $\Omega^-$ is close to 3 in both the default and string-melting versions. This suggests that apart from the $\Omega^-$ and the associated hadrons, these two event classes  have limited differences in terms of strangeness production. We compare the numbers of strange hadrons and anti-baryons between events with and without $\Omega^-$  in Table \ref{diff:b}. With reference to Table \ref{scenarios}, both the default and string-melting versions of AMPT show a mixture of the two $\Omega$-production scenarios, while the string-melting mode favors the first scenario.

\begin{table}[htb]
\centering
\caption{Differences in quantities between events with and without $\Omega^{-}$. }
\label{diff}       % Give a unique label
% For LaTeX tables you can use
\begin{subtable}{0.38\textwidth}\centering
    \caption{Average number of $s\bar{s}$ pairs. }
    \begin{tabular}{ll}
    \hline
   & $\Delta N_{pair(s\bar{s})}$ \\\hline
    AMPT-SM &  $3.07\pm0.03$ \\
    AMPT-default  & $3.16\pm0.03$\\
    \hline
    \end{tabular}
    \label{diff:a}
% Or use
%\vspace*{1cm}  % with the correct table height
\end{subtable}
\begin{subtable}{0.58\textwidth}\centering
    \caption{Numbers of $K$ (including both $K^{+}$ and $K^{0}$) and $\bar{B}$ (i.e., $\bar{\Lambda}$, $\bar{\Xi}$, $\bar{\Omega}$, $\bar{p}$ and $\bar{n}$). }
    \begin{tabular}{lll}
    \hline
   & $\Delta N_{K}$ & $\Delta N_{\bar{B}}$  \\\hline
    AMPT-SM & $2.44\pm0.02$ & $0.119\pm0.006$ \\
    AMPT-default & $1.76\pm0.03$ & $0.28\pm0.01$ \\
    \hline
    \end{tabular}
    \label{diff:b}
% Or use
%\vspace*{1cm}  % with the correct table height
\end{subtable}
\end{table}

\section{Hadron-$\Omega$ Correlation}
The results in Table~\ref{diff:b} indicate that the string-melting version of AMPT should show a stronger $K^{+}$-$\Omega^{-}$ correlation or a weaker $\bar{B}$-$\Omega^{-}$ correlation. Also, since $\bar{\Omega}^{+}$ cannot carry baryon numbers, a difference in the two ``opposite-strangeness-sign'' correlations such as $K^{+}$-$\Omega^{-}$ and $K^{-}$-$\bar{\Omega}^{+}$) may relate the $\Omega^-$ production to baryon number transport. The dynamical information about this process could be revealed by the shape of these correlations. 

\begin{figure}[htb]
\centering
\begin{minipage}[c]{0.9\textwidth}
\includegraphics[width=\textwidth]{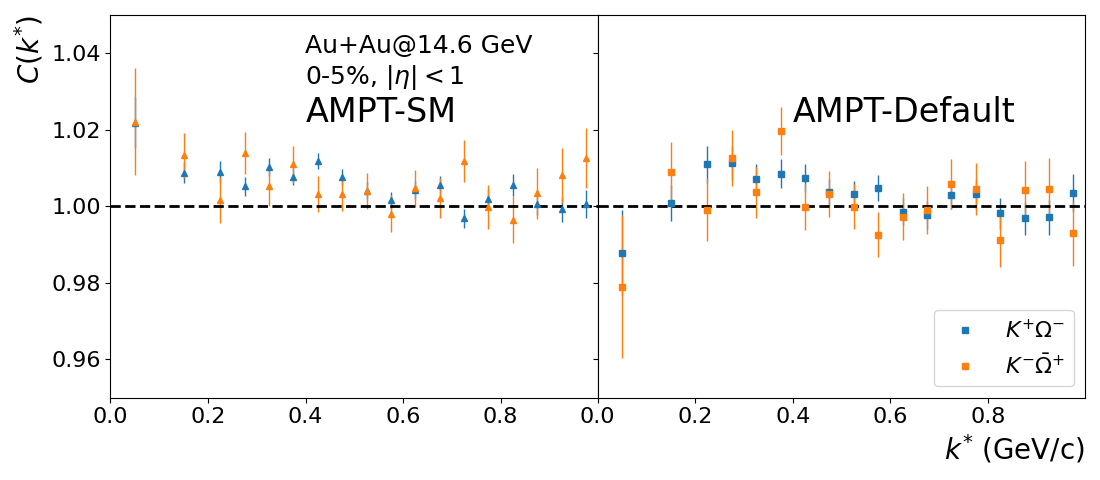}
\end{minipage}
\hspace{1cm}
\caption{AMPT simulations of ``opposite-strangeness-sign'' $K$-$\Omega$ correlation functions, $C(k^{*})=A(k^{*})/B(k^{*})$, for Au+Au collisions at 14.6 GeV.   }
\label{cf_kO}
\end{figure}

We first use the traditional event-mixing normalization to study the $k^{*}$ correlations for $K^{\pm}$-$\Omega$, $\Lambda$-$\Omega$ and $\Xi$-$\Omega$, where $k^{*}$ is half the difference between the pair-rest-frame momenta. The two-particle $k^{*}$ distribution $A(k^{*})$ is divided by the mixed-event distribution $B(k^{*})$ after normalizing each distribution at a common $k^{*}$ interval. For all hadron-$\Omega$ correlations under study, there seems to be no significant difference between the two "opposite-strangeness-sign" pair correlations with the current statistics, in either the string-melting or the default version of AMPT. An example of  $K$-$\Omega$ correlations is shown in Fig.~\ref{cf_kO}.
A potential discrepancy may exist in
the first points between the two AMPT modes, but its significance is limited by statistics.

\begin{figure}[htb]
\centering
\begin{minipage}[c]{0.9\textwidth}
\includegraphics[width=\textwidth]{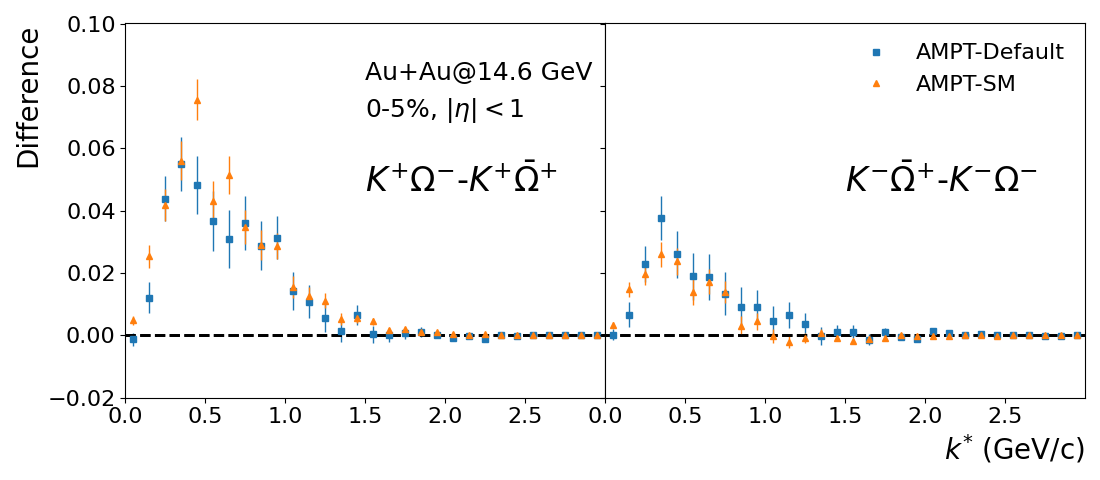}
\end{minipage}
\hspace{1cm}
\caption{
AMPT calculations of two cases of $K$-$\Omega$ correlation differences in the $k^{*}$ space: (left) $K^+$-$\Omega^- -K^+$-${\bar\Omega}^+$ and (right) $K^-$-${\bar\Omega}^+ -K^-$-$\Omega^-$, for Au+Au collisions at 14.6 GeV. }
\label{diff_k_kO}
\end{figure}

\begin{figure}[htb]
\centering
\begin{minipage}[c]{0.9\textwidth}
\includegraphics[width=\textwidth]{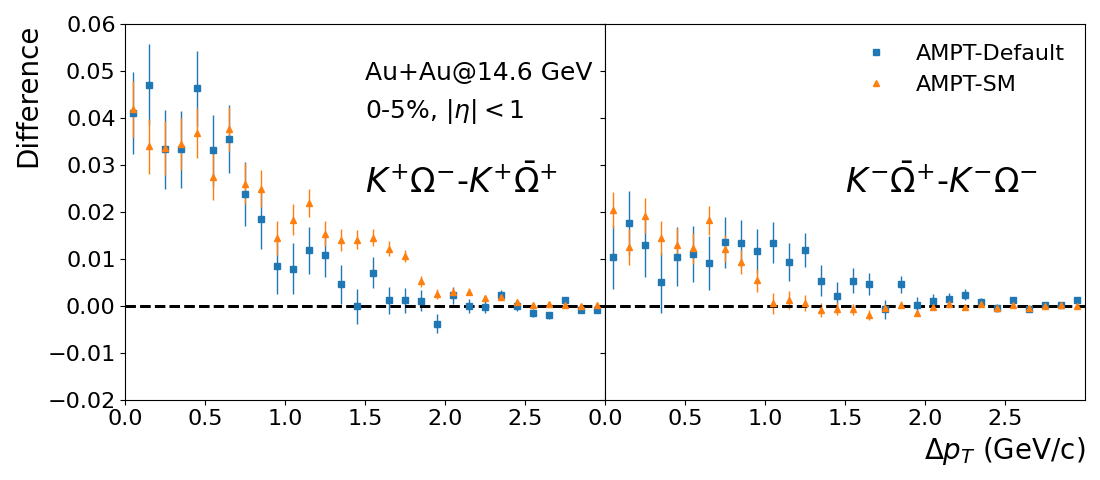}
\end{minipage}
\hspace{1cm}
\caption{AMPT calculations of two cases of $K$-$\Omega$ correlation differences in the $\Delta p_T$ space: (left) $K^+$-$\Omega^- -K^+$-${\bar\Omega}^+$ and (right) $K^-$-${\bar\Omega}^+ -K^-$-$\Omega^-$ , for Au+Au collisions at 14.6 GeV. }
\label{diff_pt_kO}
\end{figure}

We also examine the correlation  difference between the "opposite-strangeness-sign" and the "same-strangeness-sign"  distributions, each normalized by the number of events. In this definition, the latter is used as a baseline for the former so that they can be compared on the same footing and not affected by the difference between $K^{+}$($\bar{B}$) and $K^{-}$($B$) multiplicities. These  correlation differences reflect the strength and dynamic of the enhancement of $\bar{s}$($s$)-carrying hadrons in the presence of an $\Omega^{-}$($\bar{\Omega}^{+}$). In the context of the $\Omega$ production, these two differences reveal the quantitative effects of SC+BNT and SC only, respectively. As shown in Fig.~\ref{diff_k_kO} and Fig.~\ref{diff_pt_kO}, while both the string-melting and default versions of AMPT show transported-quark effects, the spreads of such effects are different in global correlation spaces, such as $\Delta p_{T}$. 

\section{Conclusion}
$\Omega$ hyperons can serve as a viable probe to various hadronization and baryon number transport mechanisms, as quark level correlations of $s\bar{s}$ may show different features in $K^{+}$($\bar{B}$)-$\Omega^{-}$ and $K^{-}$($B$)-$\bar{\Omega}^{+}$ correlation due to the presence of baryon number transport dynamics. In these proceedings, we explore two $\Omega$ production scenarios through $K$-$\Omega$ correlations using AMPT events of Au+Au collisions at $\sqrt{s_{NN}}=14.6$ GeV. With the traditional event-mixing normalization, we find no significant difference between $K^{+}$-$\Omega^{-}$ and $K^{-}$-$\bar{\Omega}^{+}$ correlation functions $C(k^{*})$ with the current statistics, which suggests that $C(k^{*})$ may be dominated by effects such as strangeness conservation or strong interaction that are common to both pairs. The correlation difference reveals transported-quark effects and a potential difference in global correlation widths between the two AMPT versions. Further investigation is needed to discern whether such differences arise from different hadronization procedures.

\begin{acknowledgement}
This work is supported by the U.S. DOE under the grant number DE-FG02-88ER40424, the National Science Foundation under Grant No. PHY-2012947, and the Natural Science Foundation of China under the grant number 11835002.
\end{acknowledgement}
%
% BibTeX or Biber users please use (the style is already called in the class, ensure that the "woc.bst" style is in your local directory)
% \bibliography{name or your bibliography database}

\end{document}